\documentclass[prb,twocolumn,showpacs,preprintnumbers,amsmath,amssymb]{revtex4}
\usepackage{graphicx}
\usepackage{dcolumn}
\usepackage{bm}

\begin{document}

\title{Magnetic phase diagram of the semi-Heusler alloys from first-principles}

\author{E. \c Sa\c s\i o\u{g}lu$^{1,2}$}\email{e.sasioglu@fz-juelich.de}
\author{L. M.  Sandratskii$^{1}$}\email{lsandr@mpi-halle.de}
\author{P. Bruno$^{1}$}\email{bruno@mpi-halle.de}

\affiliation{$^{1}$Max-Planck-Institut f\"ur Mikrostrukturphysik,
D-06120 Halle, Germany\\
$^{2}$Institut f\"ur Festk\"orperforschung, Forschungszentrum
J\"ulich, D-52425 J\"ulich, Germany}

\date{\today}

\begin{abstract}
The magnetic phase diagram of the Mn-based semi-Heusler alloys is
determined at $\textrm{T}=0$ using first-principles calculations
in conjunction with the frozen-magnon approximation. We show that
the magnetism in these systems strongly depends on the number of
conduction electrons, their spin polarization and the position of
the unoccupied Mn 3\textit{d} states with respect to Fermi energy.
Various magnetic phases are obtained depending on these
characteristics. The conditions leading to diverse magnetic
behavior are identified. We find that in the case of a large
conduction electron spin polarization and the unoccupied Mn
3\textit{d} states lying far above the Fermi level, an RKKY-like
ferromagnetic interaction is dominating. On the other hand, the
antiferromagnetic superexchange becomes important in the presence
of large peaks of the unoccupied Mn 3\textit{d} states lying close
to the Fermi energy. The overall magnetic behavior depends on the
competition of these two exchange mechanisms. The obtained results
are in very good agreement with the available experimental data.
\end{abstract}


\maketitle

Recent rapid development of spin electronics intensified the
search on the ferromagnetic materials with tunable physical
characteristics.  The Heusler alloys form a particularly
interesting class of systems. These systems have crystal
structures  and lattice parameters similar to many compound
semiconductors, high Curie temperatures and high spin polarization
at the Fermi level. Some of them were found to have a
half-metallic ground state, which is characterized by a 100\%
carrier spin polarization.\cite{Groot} These properties make the
Heusler alloys particularly attractive for use in spintronics
devices. Further prominent physical properties of this class of
materials are the martensitic transformations.\cite{LB} At low
temperatures several Heusler compounds (i.e., Ni$_2$MnGa,
Co$_2$NbSn) undergo a structural transformation from a highly
symmetric cubic austenitic phase to a low symmetry martensitic
phase giving rise to two unique effects: magnetic shape memory
effect and inverse magneto-caloric effect. Both features are
regarded promising for the development of smart materials for
future technological applications. An interesting combination of
physical properties makes Heusler alloys the subject of intensive
experimental and theoretical investigations.
\cite{GalanakisHalf,interface,Rusz,Kurtulus,Picozzi,Zayak,Takeuchi,Acet}

To predict new Heusler compounds with desired magnetic properties
the theoretical understanding of the exchange interactions in
these systems plays  an important role. Our previous studies on
experimentally well established Ni-based compounds Ni$_2$MnZ
($\textrm{Z}= \textrm{Ga}$, In, Sn, Sb) revealed a complex
character of the magnetism in these systems. In particular, the
obtained long range and oscillatory behavior of the exchange
interactions as well as their strong dependence on the \textit{sp}
atom (Z) gave an evidence for the conduction electron mediated
exchange mechanism in Heusler alloys.\cite{Sasioglu} The situation
is not so different from the experimental point of view. Early
measurements by Webster \textit{et al.} on quaternary Heusler
alloys Pd$_2$MnIn$_{1-x}$Sn$_{x}$ and Pd$_2$MnSn$_{1-x}$Sb$_{x}$
\cite{LB} and recent studies on Mn-based semi Heusler compounds
Ni$_{1-x}$Cu$_{x}$MnSb  and AuMnSn$_{1-x}$Sb$_{x}$ demonstrated
the importance of the \textit{sp}-electrons in establishing
magnetic properties.\cite{NiCuMnSb,AuMnSnSb} In particular, the
magnetic order and the Curie temperature in latter compounds  is
found to be quite sensitive to the \textit{sp}-electron number.

\begin{figure}[!b]
\includegraphics[scale=0.52]{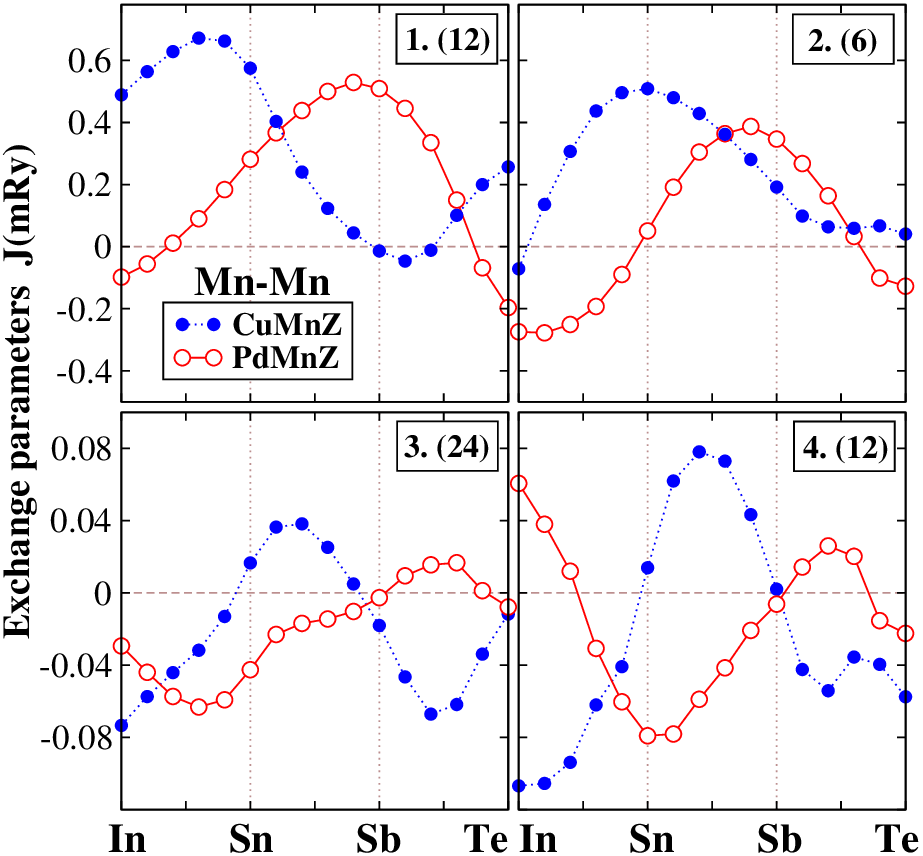}
\vspace*{-0.1cm} \caption{ (Color online) Exchange interactions
between first four nearest-neighbors of the Mn atoms in PdMnZ and
CuMnZ as a function of the \textit{sp}-electron number of the Z
constituent. Number of atoms within corresponding coordination
spheres are also given.} \label{fig1}
\end{figure}

The complexity of the exchange interactions complicates
theoretical design of the systems with given magnetic
characteristics and calls for further studies directed to find
parameters of the systems governing magnetic properties and
allowing their targeted change. Pursuing this aim we performed
systematic calculations on semi-Heusler alloys CuMnZ and PdMnZ for
different Z constituents (In, Sn, Sb, Te and their mixture treated
in the virtual crystal approximation) focusing on the
microscopical mechanisms of the formation of the long range
magnetic order. We determine the magnetic phase diagram of these
systems at zero temperature and estimate the $T_{\textmd{C}}$ of
the compounds having ferromagnetic ground state within mean-field
approximation. The calculational method is described elsewhere in
detail.\cite{Sasioglu} We use the lattice constants of  $6.26 \AA$
and $6.09 \AA$ for PdMnZ and CuMnZ, respectively.\cite{LB}

\begin{figure}[t]
\begin{center}
\includegraphics[scale=0.52]{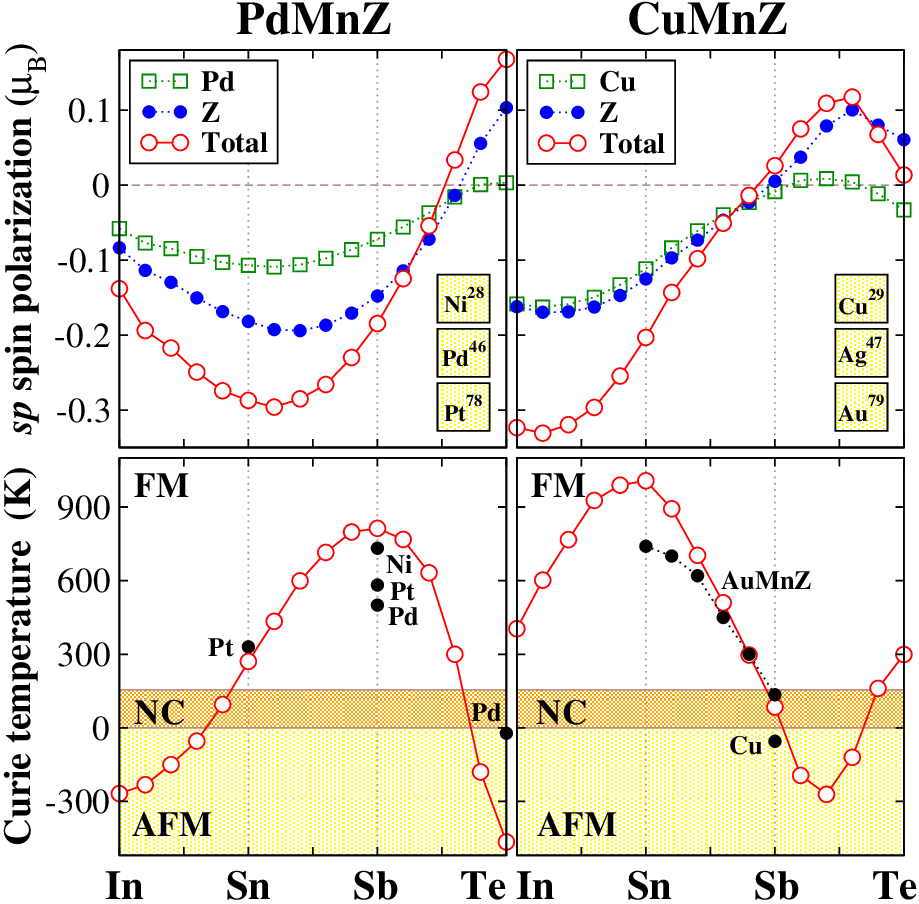}
\end{center}
\vspace*{-0.4cm} \caption{ (Color online) Upper panels: The
\textit{sp}-electron spin polarization of the X=Pd, Cu and Z
constituents. Lower panels: Ground state magnetic phase diagram
and mean-field Curie temperature ($T^{\textmd{MFA}}_\textmd{C}\sim
J_0\equiv\sum_{R \ne 0}J_{0R}$) of PdMnZ and CuMnZ. FM, NC and AFM
stand for ferromagnetic, non-collinear, and antiferromagnetic
ordering, respectively . Experimental values of the Curie
temperatures (filled spheres) are taken from
Refs.~\onlinecite{LB,AuMnSnSb}.} \label{fig2}
\end{figure}

We begin with the discussion of the exchange interactions
calculated within the frozen-magnon approximation \cite{Magnon}.
Fig.~\ref{fig1} presents the obtained exchange parameters between
first four nearest-neighbors of the Mn atoms in PdMnZ and CuMnZ as
a function of the \textit{sp}-electron number of the Z
constituent. The corresponding Curie temperatures are given in
Fig.~\ref{fig2}. In agreement with the experiments and the results
of our previous calculations we obtain a strong dependence of the
exchange parameters  on the Z constituent. As it is seen from
Fig.~\ref{fig1} all exchange parameters oscillate between
ferromagnetic and antiferromagnetic values with the variation of
the \textit{sp}-electron concentration. Considering first two
nearest neighbor exchange parameters  we see that they have
ferromagnetic character for a broad range of compositions and
dominate over the rest parameters playing a decisive role in
determining the magnetic phase diagram (see Fig.~\ref{fig2}). The
absolute value of the parameters (we present only four nearest
neighbor parameters) decays quickly with increasing interatomic
distance. The parameters show the RKKY-type oscillations.

This complex behavior of the exchange interactions appeared as a
rich spectrum in the magnetic phase diagram (see Fig.~\ref{fig2})
where magnetic order changes with \textit{sp}-electron
concentration. Furthermore, an interesting observation for the
magnetism in these systems is that the maximum of the exchange
interactions for both PdMnZ and CuMnZ corresponds to the similar
number of the \textit{sp}-electrons. The shift of the maxima for
two systems is explained by the fact that Pd has one
\textit{sp}-electron less than Cu. The properties of the exchange
interactions are reflected in the properties of the Curie
temperature (see Fig.~\ref{fig2}) where we also obtained a
relative shift of the maxima of the two curves corresponding to
one \textit{sp}-electron.

Combination of these features with the large distance between Mn
atoms ($d_{Mn-Mn}>4$ \AA) makes the assumption natural that the
exchange coupling mechanism in Heusler alloys is indirect and
takes place via the conduction electrons. A RKKY-type coupling
seems to be the most probable one, however, the rich spectrum of
magnetic orderings obtained by varying \textit{sp}-electron
concentration (Fig.~\ref{fig2}) makes the interpretation of the
results in terms of existing theories difficult and calls for a
detailed analysis. This is explained by the fact that the DFT is
not based on a model Hamiltonian approach and does not use a
perturbative treatment. Thus, various exchange mechanisms appear
in the calculational results in a mixed form and a separation of
them into individual contributions is not easy. In this case the
model Hamiltonian studies relevant to the problem provide useful
information for a qualitative interpretation of the DFT results.
Among such approaches the Anderson \textit{s-d}
model\cite{Anderson} is the most suitable for Heusler alloys
because of the localized nature of the magnetism in these
systems.\cite{LB} Within this model the effective exchange
interaction between distant magnetic atoms can be separated into
two contributions:
$J_{\textrm{indirect}}=J_{\textrm{RKKY}}+J_{\textrm{S}}$. The
first term is an RKKY-like ferromagnetic term which stems from the
interaction between the local moment and the conduction electron
states inducing a spin polarization in conduction electron sea. In
general, this interaction contains two distinct processes:
electrostatic Coulomb exchange interaction and \textit{sp-d}
mixing interaction. The former induce a net positive spin
polarization while the contribution of latter is always negative
and disappears in strong magnetic limit.\cite{Watson} Note that
the amplitude of this polarization determines the strength of the
RKKY-like exchange coupling. However, the second term has
different origin, it arises from virtual-charge excitations in
which electrons from local \textit{d} states of the Mn are
promoted above the Fermi sea providing an additional contribution
to the indirect exchange coupling. This term depends mostly on the
distance of the unoccupied Mn 3\textit{d} peaks from the Fermi
level. The closer the peaks to the Fermi level the stronger the
$J_{\textrm{S}}$. Moreover, in contrast to the first term, this
second term is always antiferromagnetic and its strength decays
exponentially with distance. The overall magnetic behavior depends
on competition of these two terms as exemplified by the present
systems.

\begin{figure}[t]
\begin{center}
\includegraphics[scale=0.52]{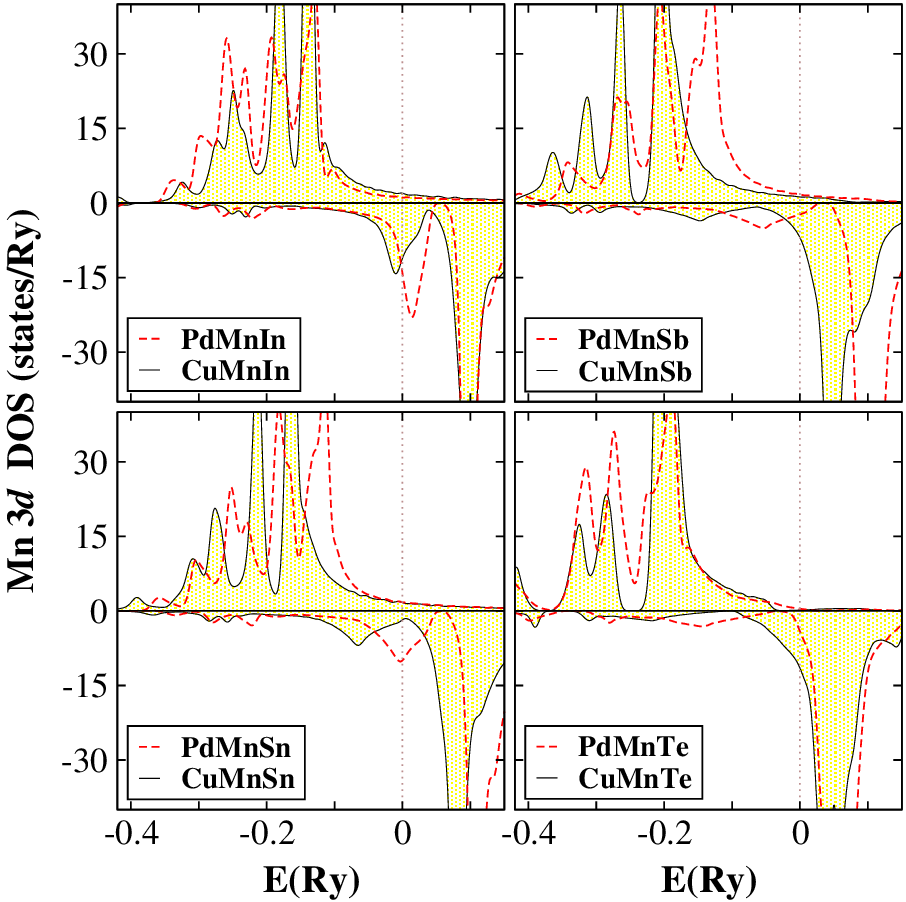}
\end{center}
\vspace*{-0.5cm} \caption{(Color online) Spin resolved Mn
3\textit{d} density of states in PdMnZ and CuMnZ
($\textrm{Z}=\textrm{In}$, Sn, Sb, Te) for stoichiometric
compositions. Vertical dotted lines denote the Fermi level.}
\label{fig3}
\end{figure}

Now we turn to the interpretation of the DFT results in terms
of these two mechanisms. To estimate the relative contribution of
the ferromagnetic RKKY-like exchange we present in Fig.~\ref{fig2}
the calculated conduction electron spin polarization in PdMnZ and
CuMnZ as a function of the electron number of the Z constituent.
The analysis of the calculational data allows to make a number of
important conclusions. First, in both systems the direction of the
induced spin polarization is opposite to the direction of the Mn
moment in a broad interval of compositions. This feature reveal
the primary role of underlying \textit{sp-d} mixing interaction in
exchange coupling and justifies the use of the Anderson
\textit{s-d} model for the description of the magnetism in these
systems. It is worth to note that, in agreement with the present
findings, recent experiments on ferromagnetic full-Heusler
compounds Cu$_2$MnAl and Ni$_2$MnSn gave a large conduction
electron spin polarization that is antiferromagnetically coupled
to Mn magnetic moment.\cite{polarization_1,polarization_2}

Another remarkable feature we  obtained is a very clear relation
between the spin polarization and the mean-field Curie temperature
(or exchange parameters) in a large part of the phase diagram (see
Fig.~\ref{fig2}). Indeed, the compounds with very large spin
polarization are characterized by very high Curie temperatures.
Interestingly, for the zero polarization also the Curie
temperature vanishes reflecting the dominating character of the
ferromagnetic RKKY-like exchange mechanism in establishing
magnetic properties. However, at some regions superexchange
mechanism becomes important. This can be seen in Fig.~\ref{fig2}
where for PdMnIn$_{1-x}$Sn$_{x}$ ($x< 0.8$) system we obtain an
antiferromagnetic order in spite of very large spin polarization.
To gain further insight into the nature of this coupling we
present in Fig.~\ref{fig3} the spin resolved Mn 3\textit{d}
density of states of PdMnZ and CuMnZ for stoichiometric
compositions. As pointed out earlier this mechanism is more
sensitive to the properties of the unoccupied Mn 3\textit{d}
peaks. For PdMnIn there is a large peak just above the Fermi level
and, as a result the antiferromagnetic superexchange dominates
over the ferromagnetic RKKY exchange giving rise to an AFM order.
In transition from In to Sn this peak gradually decreases and,
therefore, the superexchange becomes less important. Around
$\textrm{Z}=\textrm{Sb}$ this peak almost disappears leading to
the FM order. However, another large peak becomes close to the
Fermi level when $\textrm{Z}$ approaches $\textrm{Te}$. This turns
the system from a ferromagnetic state into a antiferromagnetic one
with a non-collinear ordering in between due to competition of
these two mechanisms. The situation is very similar in the case of
CuMnZ where we also obtain a rich magnetic behavior.

Finally, we would like to comment on the magnetic behavior of the
systems having non-magnetic X or Z elements from the same column
of the Periodic table ($\textrm{X}=\textrm{Ni}$, Pd, Pt or Cu, Ag,
Au, $\textrm{Z}=\textrm{Al}$, Ga, In, etc). For these non-studied
compounds we expect a similar behavior on the basis of previous
experimental and theoretical  studies.\cite{LB,Sasioglu} Indeed,
as seen in Fig.~\ref{fig2} the XMnSb compounds with
$\textrm{X}=\textrm{Ni}$, Pd, Pt have similar values of the Curie
temperature and are strongly ferromagnetic. Our results for ground
state magnetic order and finite temperature properties compare
well with the existing experimental data. In particular, the
observed trend in the variation of the Curie temperature of
AuMnSn$_{1-x}$Sb$_{x}$ is very well described by our
calculations.\cite{AuMnSnSb}

In conclusion, the magnetic behavior of the semi-Heusler alloys
can be described in terms of the competition of two exchange
mechanisms: ferromagnetic RKKY-like exchange and antiferromagnetic
superexchange. We found that each mechanism depends on certain
parameter. In the case of large conduction electron spin
polarization and the unoccupied Mn 3\textit{d} states lying far
from the Fermi level, an RKKY-like ferromagnetic interaction is
dominating while antiferromagnetic superexchange becomes important
in the presence of the peaks of unoccupied Mn 3\textit{d} states
close to the Fermi energy. Various magnetic phases are obtained
depending on these characteristics. These findings suggest a
practical tool for the design of the materials with given
properties.

%-----------------------------------------------------------------

\end{document}